# An Improved Model for Relativistic Solar Proton Acceleration applied to the 2005 January 20 and Earlier Events

D. J. Bombardieri[1], M. L. Duldig[2], J. E. Humble[3,2] and K. J. Michael[1,4]

[1]*Institute for Antarctic and Southern Ocean Studies, University of Tasmania, Hobart, Tasmania, Australia 7001*
[2]*Australian Antarctic Division, Kingston, Tasmania, Australia 7050*
[3]*School of Mathematics and Physics, University of Tasmania, Hobart, Tasmania, Australia 7001*
[4]*Antarctic Climate and Ecosystems Cooperative Research Centre, University of Tasmania, Hobart, Tasmania, Australia 7001*

e-mail of corresponding author:  marc.duldig@aad.gov.au

ABSTRACT

This paper presents results on modelling the ground level response of the higher energy protons for the 2005 January 20 ground level enhancement (GLE). This event, known as GLE 69, produced the highest intensity of relativistic solar particles since the famous event on 1956 February 23. The location of recent X-ray and γ-ray emission (N14° W61°) was near to Sun-Earth connecting magnetic field lines, thus providing the opportunity to directly observe the acceleration source from Earth. We restrict our analysis to protons of energy ≥450 MeV to avoid complications arising from transport processes that can affect the propagation of low energy protons. In light of this revised approach we have reinvestigated two previous GLEs: those of 2000 July 14 (GLE 59) and 2001 April 15 (GLE 60). Within the limitations of the spectral forms employed, we find that from the peak (06:55 UT) to the decline (07:30 UT) phases of GLE 69, neutron monitor observations from 450 MeV to 10 GeV are best fitted by the Gallegos-Cruz & Perez-Peraza stochastic acceleration model. In contrast, the Ellison & Ramaty spectra did not fit the neutron monitor observations as well. This result suggests that for GLE 69, a stochastic process cannot be discounted as a mechanism for relativistic particle acceleration, particularly during the initial stages of this solar event. For GLE 59 we find evidence that more than one acceleration mechanism was present, consistent with both shock and stochastic acceleration processes dominating at different times of the event. For GLE 60 we find that Ellison & Ramaty spectra better represent the neutron monitor observations compared to stochastic acceleration spectra. The results for GLEs 59 and 60 are in agreement with our previous work.

INTRODUCTION

Relativistic protons produced from large solar eruptive episodes represent a direct sample of matter from some of the most energetic processes in the solar system (e.g., solar flares and coronal mass ejections (CMEs)). These processes can convert in excess of $10^{32}$ ergs of magnetic energy into kinetic (accelerated particles) and thermal (heated plasma) energies on timescales of a fraction of a second to several tens of minutes. Energetic particles which escape the solar corona are guided via the interplanetary magnetic field (IMF) and, as they approach the Earth, are deflected by the Earth's magnetic field. For a sea-level neutron monitor to detect the secondary neutrons produced in an atmospheric cascade, the primary particle energy must exceed 450 MeV.

An important question in solar physics concerns the mechanisms responsible for the production of the relativistic protons that give rise to ground level enhancements (GLEs) in the cosmic ray flux. In the past, solar energetic particle events (SEPs) (which include GLEs) were divided as either impulsive or gradual. In impulsive events, particle acceleration was attributed to flare processes such as resonant wave-particle interactions following magnetic reconnection. Reames (1999) described impulsive events as originating from a narrow range of solar longitudes that are magnetically well connected to the observer, whilst in gradual events particles were accelerated over a broad range of heliolongitudes by CME-driven shocks. GLEs are characteristically associated with both X-class solar flares and CMEs, making it difficult to isolate the key particle acceleration signatures.

Recent studies have shown that the impulsive and gradual classifications represent the extremes of a combination of processes. Tylka et al. (2005) propose that the seed population is likely to be a combination of flare and supra-thermal solar wind particles which are then accelerated by CME-driven shocks. Conversely, Cane et al. (2007) point out that two processes are likely to co-exist, one related to solar flares and the other to CME shocks. The evolution of any event depends on the relative importance of the two processes. Although most investigators now agree that more than one process is present in SEP generation the details of the processes themselves is still a source of debate.

This paper is the last in a series of three works investigating the mechanism/s responsible for the GLEs of 2000 July 14 (GLE 59); 2001 April 15 (GLE 60) and 2005 January 20 (GLE 69). Previous investigations by Bombardieri et al. (2006) (hereafter Paper 1) and Bombardieri et al. (2007) (hereafter Paper 2) have shown that CME-driven shocks and processes associated with magnetic reconnection (e.g., resonant wave-particle interactions) are important in accelerating the relativistic protons which gave rise to GLEs 59 and 60.

For this study we investigate the 2005 January 20 solar event (GLE 69), which produced the highest intensity of relativistic solar particles since the famous event on 1956 February 23 (GLE 05). The largest 1-minute record at Terre Adélie was ~46 times the pre-event level. This is the largest increase seen at sea-level since 1956 (Leeds ~47 times in 15-minute data). In addition, intense X-ray and γ-ray emission (N14° W61°, http://helios.izmiran.rssi.ru/cosray/events05.htm) was observed near to the nominal Sun-Earth connecting magnetic field lines, allowing for direct observations of the acceleration source from Earth.

The aim of this study is to gain insight into the acceleration process/es responsible for the production of relativistic protons that gave rise to GLE 69. The energy spectra of relativistic protons carry information about the acceleration process, providing a useful tool for probing the source mechanism. We fit analytical and numerical functions representing shock and stochastic acceleration mechanisms to neutron monitor observations up to 10 GeV. In § 2 we discuss satellite and ground-based observations of GLE 69. In § 3 we give a brief description of the global analysis technique used to model the arrival of relativistic particles at 1 AU and present the results. In § 4 we discuss a revised approach to our modelling to account for transport processes which could affect the propagation of low-energy protons and present the results. In § 5 we briefly discuss particle transport conditions during GLE 69 and consider the source mechanisms which may have led to relativistic proton production for this event. In addition, we re-investigate the relativistic proton acceleration processes which gave rise to GLEs 59 and 60 and compare and contrast these new results with our previous investigations. In § 6 we present our conclusions.

2. OBSERVATIONS

The 2005 January 20 solar event was associated with a GOES classified X7.1/2B solar flare and fast CME. This event represented the largest of a series of solar eruptions that occurred during a period of intense solar activity extending from 14 to 20 January. The primary source of this activity was NOAA active region 10720. The X-ray flare began at 06:36 UT and peaked at 07:01 UT. The CME was first detected by the LASCO instrument on board SOHO at 06:54 UT. Simnett (2006) estimated CME onset and speed at approximately 06:40 UT and 2500 km s$^{-1}$ respectively. RHESSI (in Earth orbit) observed a peak in γ-ray emissions in the 4 to 7 MeV energy region at approximately 06:46 UT (Simnett 2006). The SONG instrument on board CORONAS-F observed γ-ray emissions with a time profile similar to RHESSI. Kuznetsov et al. (2005) suggest that protons with energies >300 MeV were accelerated at the Sun between 06:38:30 ST and 06:42:30 ST (where "ST" refers to Solar Time, the Universal Time (UT) of an event at the sun).

GLE onset commenced at 06:48 UT in 1-minute neutron monitor data. The event had a rapid rise, peaking within 5 minutes at many stations, followed by a decay lasting many hours. At 14:00 UT the increase above the galactic cosmic ray background was still ~10% at Terre Adélie. The largest neutron monitor response was observed at the high-altitude station South Pole with a maximum in 1-minute data of ~ 5440 % above the pre-increase level. The rapid rise (e.g., 5 minutes) seen in the Terre Adélie, McMurdo and South Pole neutron monitor intensity/time profiles (Figs. 1a, 1b and 1c) indicates that relativistic protons had rapid access to Sun-Earth connecting field lines. Other neutron monitors recorded different responses (e.g. Figs 1d, 1e and 1f).

The Erevan neutron monitor (cut-off 7.6 GV, altitude 3200 m) recorded a small but significant increase. Small increases were also seen by the Tibet neutron monitor (cut-off 14.1 GV, altitude 4300 m) (Myasaki et al. 2005), the GRAND muon telescope (D'Andrea and Poirer 2005) and the Aragats muon detector (Bostanjyan et al. 2007), indicating the presence of very low fluxes of particles >15 GeV.

Our modelling of the sea-level neutron monitor response incorporates the Debrunner et al. (1982) yield function. Application of this function indicates that the low fluxes of higher rigidity particles observed in muons and at Tibet will not produce significant responses at other neutron monitors with high cut-offs. This is consistent with our results for spectra, pitch angle distributions and arrival directions.

Corrections of observed increases to a standard sea-level atmospheric depth of 1033 g cm$^{-2}$ were made using the two-attenuation length method of McCracken (1962). An attenuation length of 100 g cm$^{-2}$ was derived from a comparison of Mt. Washington and Durham neutron monitors. The results presented in the following section are relatively insensitive to changes in the attenuation length (e.g., 90, 95 and 110 g cm$^{-2}$). In order to calculate the absolute flux of solar particles, it is necessary to select a low altitude station with one of the largest increases as the normalization station (Cramp 1996). After correcting observed increases to standard sea-level atmospheric depth, Terre Adélie was found to have the largest response (Fig. 1a) and was used as the normalization station for this analysis. Figure 2 shows the viewing directions (in geographic coordinates) of selected neutron monitors at 06:55 UT (peak phase). ACE measurements indicate that the IMF direction was located at high southern latitudes (as indicated by the large black circle in Fig. 2). This explains the substantial increases observed at Terre Adélie, McMurdo and South Pole stations (Figs. 1a, 1b and 1c), whilst the very much smaller increases observed at SANAE and Mawson (Figs. 1d and 1e respectively) indicate pitch angle distributions that are both extremely anisotropic and asymmetric.

3.0. MODELLING THE NEUTRON MONITOR RESPONSE

The global method for modelling the solar cosmic ray ground level response by neutron monitors has been developed over many years (Shea & Smart 1982; Humble et al. 1991) and is described in detail in Cramp et al. (1997a) and Paper 1. The technique derives the spectrum, the axis of symmetry of the particle arrival and anisotropy of relativistic solar protons that give rise to the increased neutron monitor response. The modelling procedure employs a least-squares method to efficiently analyse parameter space for optimum solutions. The geomagnetic field model of Tsyganenko (1989), with IGRF 2005 parameters and adjustments for Kp and the Dst index, was employed to determine the asymptotic viewing directions of ground-based instruments (Flückiger & Kobel 1990; Boberg et al. 1995).

3.1. *Results*

Observations from 41 neutron monitors (Table 1) were modelled every five minutes between 06:50 and 08:30 UT. Figure 3 shows the observed increases, corrected to standard sea-level pressure, at selected neutron monitor stations and the model fits to the observations. Good fits to observations were achieved for most stations at the peak and decline phases of the event. However, during the rising phase of the event the model significantly underestimated the response at South Pole and significantly overestimated the response at McMurdo and Mawson. We believe this is due to the difficulty the model had in accounting for the extreme asymmetric anisotropy of the event during this interval. For example, model fits are sometimes poor for stations with viewing cones located near the edge of the steep part of the pitch angle distribution affected by GLEs. This is due to the coarse nature of the neutron monitor cell size, combined with differences between the real, dynamic, geomagnetic field, (which can vary from minute to minute), and the Tsyganenko geomagnetic field model with corrections for Kp (3-hourly) and Dst (hourly) variations. Any timing uncertainties during the rapidly rising phase can also compound the problem, particularly in very large impulsive events.

3.2. *Arrival directions*

Figure 4 illustrates the GSE longitude and latitude of the axis of symmetry of the particle pitch angle distribution ("arrival direction"), together with the IMF direction as measured by ACE at the Earth-Sun Lagrangian point L1. The average GSE longitude of the IMF direction as measured by ACE during the first 30 minutes of the event was approximately -60º. At 06:55 and 07:00 UT we find the apparent arrival longitude to be in good agreement with the measured field longitude. However, there is inconsistency between the apparent arrival longitude and the measured longitude of the field direction from 07:05 UT onwards. By 07:30 UT the model longitude is approximately 60º east of the measured field longitude.

From 06:55 to 07:15 UT the latitude of the arrival direction is in good agreement with the measured field latitude (i.e., centred at high southern latitudes in GSE coordinates). However, from 07:15 UT the apparent arrival latitude began to move north, and by 07:30 UT was approximately 45º north of the measured field latitude. It is important to note that the magnetic field measured at a point is not necessarily the same as the average field sampled by the particle over its orbit. For example, a 2 GV proton in a typical interplanetary field has a Larmor radius of ~0.01 AU, of the order of the coherence length of interplanetary magnetic turbulence. Therefore, model flow vectors need not necessarily align with the measured magnetic field vector (Bieber et al. 2002).

We estimate the uncertainty for the particle "arrival directions" at 06:55 UT to be ±1° in latitude and ±18° in longitude. At 08:00 UT these uncertainties are estimated to be ±10° in latitude and ±21° in longitude. Uncertainties for parameters at most other solutions will lie between these values.

3.3. *Pitch Angle Distributions*

The particle pitch angle α is defined as the angle between the particle velocity and the mean magnetic field. The distribution used is a simplification of the exponential form described by Beeck & Wibberenz (1986) and is presented in Cramp et al. (1997b). It has the functional form

$$G(\alpha) = exp\left[\frac{-0.5(\alpha - \sin\alpha \cos\alpha)}{A - 0.5(A - B)(1 - \cos\alpha)}\right] \quad (1)$$

where *A* and *B* are variable parameters (Cramp et al. 1997a). Parameter A has most effect on the width of the anisotropy while B has most effect on the relative flux at pitch angles > 90°. The distribution can be considered as having an anisotropic component (representing particles which arrive directly from the Sun) and an isotropic component (where the effects of local scattering dominate the distribution).

The temporal development of the pitch angle distribution between 06:50 and 08:00 UT is illustrated in Figure 5 and the fitted parameters along with their uncertainties are listed in Table 2. The particle arrival at the peak (06:55 UT) was strongly anisotropic (Fig. 5a). This suggests that during the initial phase of the event, relativistic protons injected into the interplanetary medium arrived directly from the Sun without being affected by local scattering. However, the degree of anisotropy decreased (Fig. 5b) and by 08:00 UT, well into the decline phase, there is evidence for significant scattering (Fig. 5c).

3.4. *Spectrum*

The spectral form used in this analysis is a modified power law in rigidity (equation 2), as described in Cramp et al. (1997b).

$$J_{\parallel} = KP^{-(\gamma-\delta\gamma(P-1))} \tag{2}$$

where $J_{\parallel}$ is the peak cosmic ray flux arriving from the Sun along the axis of symmetry of the pitch angle distribution. The parameters are the particle rigidity ($P$), the parallel flux at 1 GV ($K$), the power law exponent ($\gamma$) and the change of $\gamma$ per GV ($\delta\gamma$), where a positive value of $\delta\gamma$ results in a spectrum that steepens with increasing rigidity.

Particle spectra derived between 06:50 and 07:15 UT are illustrated in Figure 6 and fitted parameters along with their uncertainties are listed in Table 3. The uncertainty in the calculated flux at 1 GV is less than 10 %.

Model fits to observations for the 06:50 to 06:55 UT interval (rise phase) had large residuals for some stations, due to the extreme asymmetric anisotropy (see section 3.1). Nonetheless, our model produced a spectral index ($\gamma$) of -5.5 with a small change of slope parameter ($\delta\gamma$) of 0.04. At 06:55 UT (peak) the spectral index was very soft at -9.2±0.6, again with a small change of slope parameter ($\delta\gamma$) of 0.07. The index hardened during the decline phase, ranging from -7.7±0.5 at 07:00 UT to -7.0±0.4 an hour later, with $\delta\gamma$ having insignificant values. These spectra agree with results reported by Bütikofer et al. (2006) and Plainaki et al. (2007).

4.0. A REVISED APPROACH

In Papers 1 and 2 we included low energy spacecraft observations of proton intensities in our spectral fits to determine the acceleration process. However, interplanetary processes such as pitch angle scattering (due to resonant interactions and small-scale magnetic inhomogeneities) and magnetic cloud structures can affect the propagation of low energy protons en-route to Earth. In addition, for major solar events where a CME-driven shock is generated, streaming particles become trapped near the shock by self-generated Alfvén waves (i.e., streaming-limited intensities), flattening the spectra of escaping particles at low energies (Reames 1999). Because the gyroradii of relativistic protons ≥450 MeV are at least equal to or greater than the coherence length of interplanetary magnetic field turbulence they are less likely to be affected by interplanetary transport effects. Therefore, to account for these effects, we fit analytical spectra only to neutron monitor energies (≥450 MeV) in order to determine the acceleration process more accurately.

4.1. *Particle Acceleration Spectra*

Particle acceleration within the solar corona may in principle occur in a variety of ways: direct particle acceleration in neutral current sheets by DC electric fields; stochastic acceleration through the process of resonant wave-particle interactions; and acceleration at coronal shocks either CME- or flare-driven. However, current theoretical models of direct particle acceleration via DC electric fields fail because they cannot explain the presence of energetic protons in the GeV energy domain (Miller et al., 1997). Therefore, we focus our attention on shock and stochastic acceleration processes, both of which are capable of accelerating protons to relativistic energies.

4.1.1 *Shock Acceleration Spectra*

Ellison & Ramaty (1985) produced equation (3) to approximate the form of diffusive shock acceleration.

$$\left(\frac{dJ}{dE}\right) \propto \left(\frac{dJ}{dE}\right)_0 \exp\left(-\frac{E}{E_0}\right) \tag{3}$$

where

$$\left(\frac{dJ}{dE}\right)_0 \propto n_{inj}\left(E_i^2 + 2E_i m_0 c^2\right)^{3/[2(r-1)]}\left(E^2 + 2E m_0 c^2\right)^{-(1/2)[(r+2)/(r-1)]}$$

(4)

and $(dJ/dE)_0$ is the differential particle intensity, particles (cm$^2$ s sr MeV)$^{-1}$

$n_{inj}$ is the number density of seed particles injected far upstream of the shock,

*c* is the speed of light,

$m_oc^2$ is the proton rest mass energy,

*E* is the particle energy in MeV,

*r* is the shock compression ratio.

In shock acceleration, particles are able to gain energy by scattering multiple times between magnetic field irregularities both upstream and downstream of the shock. The compression at the shock is the source of the energy. The exponential turnover in equation (3) was incorporated to account for the various effects that might limit the number of particles accelerated to higher energies in an actual three-dimensional shock. These effects include the finite temporal evolution of the shock compared to particle acceleration times and the finite spatial distribution of the shock compared to particle diffusion lengths.

Particle acceleration is also thought to be less effective above the energy $E_0$ (e-folding energy) when proton intensities can no longer sustain the growth of resonant waves. This process leads to the leaking of high-energy particles from the acceleration region, thereby truncating the power law behaviour (Reames 1999).

4.1.2. *Stochastic acceleration Spectra*

Stochastic acceleration is broadly defined as any process in which a particle can gain or lose energy in a short time interval, but where the particle systematically gains energy over longer intervals of time (Miller et al. 1997). Plasma instabilities can arise near strong current sheets, which are prone to collapse (dissipate). This results in magnetic reconnection, which rapidly converts stored magnetic energy into kinetic and thermal energy. High-velocity plasma outflow jets (Innes et al., 1997; Miller et al., 1997; Galsgaard et al., 2005) are produced as a result of the magnetic reconnection process. Such jets represent a potential source of magnetohydrodynamic (MHD) turbulence which can initiate stochastic acceleration.

Perez-Peraza & Gallegos-Cruz (1994) and Gallegos-Cruz & Perez-Peraza (1995) presented solutions to the Fokker-Planck equation in the energy domain. These solutions are valid over the entire energy range (i.e., non-relativistic, trans-relativistic, ultra-relativistic) for both time-dependent and steady-state conditions. A selective injection process is required to supply particles into the acceleration region where they are then re-accelerated to higher energies via a stochastic process. Equation (5) is a steady-state numerical solution to the Fokker-Planck equation and incorporates a realistic injection function which represents an initial acceleration phase by DC electric fields within a neutral current sheet (Gallegos-Cruz & Perez-Peraza 1995).

$$N(E) \cong \frac{9.44 \times 10^{-7}}{(a\alpha/3)^{1/2}} \times \frac{\varepsilon^{1/2} \exp(-J_E)}{\beta^{1/4}} \times \int_{E_0}^{E} \frac{\exp\left[J_{E'} - 1.12(E'/E_c)^{3/4}\right]}{E'^{1/4} \varepsilon'^{3/2} D^{1/4}(E')} dE' \qquad (5)$$

where

$\beta = (\varepsilon^2 - m^2 c^4)^{1/2}/\varepsilon$

$D = (a/3)\beta^3 \varepsilon^2$

$E_c = 1.7926 \times 10^3 (B^3 L/n) = 4.5$ MeV.

Here $N(E)$ is particles per unit energy; $a$ and $J$ are analytical functions of energy as described by Gallegos-Cruz & Perez-Peraza (1995) and Perez-Peraza et al. (2006); $\varepsilon$ is the energy + proton rest mass energy; $B$ is the background magnetic field strength in the neutral current sheet (5 x 10$^{-4}$ T); $n$ is the local particle number density (10$^{13}$ cm$^{-3}$); $L$ is the length of the neutral current sheet (10$^7$ cm); and $\alpha$ is the acceleration efficiency.

The analytical spectra deduced from neutron monitors were used to generate the input to the fitting routine at selected energies spaced evenly on a logarithmic scale. The fitting routine is a generalised non-linear least squares program (GNLS) with data points weighted by errors in the flux data.

## 4.2. *Results: GLE 69 Spectra*

Tables 4 and 5 present the results of spectral fits to neutron monitor data for GLE 69 from the peak (06:55 UT) to the decline (07:30 UT) phase of the event. Figure 7 illustrates the results of fitting the Ellison & Ramaty and stochastic acceleration spectra to relativistic proton fluxes determined from neutron monitor observations at 1 AU for four equally-spaced times during this period. A Kolmogorov-Smirnov test at 95% confidence shows that all post-fit residuals were random, giving confidence in the weighted sum of squares result. Furthermore, for every fit, an additional test for randomness in the residuals, in the form of a sign test, demonstrated that the residuals were randomly distributed.

Table 5 lists the results and standard errors for the parameters (normalization factor $N$ and acceleration efficiency $\alpha$) of the stochastic model. Figure 8 (07:10 UT) shows residuals plotted against particle energy, clearly illustrating the better fit of the stochastic model. For the time intervals modelled, the acceleration efficiency $\alpha$ ranged from 0.01 to 0.02 $s^{-1}$. This implies that protons with injection energy of 1 MeV need only a small acceleration efficiency to produce the observed response, consistent with values reported from previous studies (Miller et al. 1990; Miller 1991).

## 4.3. *Results: GLE 59 Spectra*

Table 6 lists the results and standard errors for the parameters (compression ratio $r$ and e-folding energy $E_0$) of the Ellison & Ramaty spectral form. These results are similar to those reported previously in Table 3 of Paper 1, which included spacecraft observations.

Table 7 lists the results and standard errors for the parameters ($N$ and $\alpha$) of the stochastic model. Once again these results are similar to those reported previously in Table 8 of Paper 2.

## 4.4. *Results: GLE 60 Spectra*

Table 8 lists the results and standard errors for the parameters ($r$ and $E_0$) of the Ellison & Ramaty spectral form. The results for intervals 14:20 and 14:30 UT are similar to those reported previously. However, the value of the compression ratio for the interval 14:45 UT is slightly lower than the value quoted in Table 3 of Paper 2 (1.83 and 1.92 respectively), as a result of not including the lower energy protons.

Table 9 lists the results and standard errors for the parameters ($N$ and $\alpha$) for the stochastic model. These results are similar to those reported in Table 5 of Paper 2.

*4.5. IMF Path Length and Injection Time Estimates*

If the path length through the IMF during a GLE is known, one can calculate the expected arrival times at the Earth of protons for any energy for a specific time interval. This allows us to investigate energy spectra at certain time intervals more accurately and make deductions about the interplanetary transport processes which may have affected low-energy proton intensities.

The time $\Delta t$ required for particles to travel along the IMF is:

$$\Delta t = s/\beta c. \qquad (6)$$

where $s$ is the distance from the Sun along a nominal IMF line to Earth and $\beta$ is the particle speed in units of the speed of light $c$.

Following Lockwood et al. (1990), the distance depends on the solar wind speed $V_{SW}$ and the angular speed $\Omega$ of the Sun and is given by equation (2):

$$s = \frac{r(1+\alpha^2 r^2)^{0.5}}{2} + \frac{\ln\{\alpha r + (1+\alpha^2 r^2)^{0.5}\}}{2\alpha} \qquad (7)$$

with $\alpha = \Omega \cos \Lambda / V_{SW}$, where $\Lambda$ is the heliographic latitude and $r$ is the heliocentric radial distance. The angular speed of the sun varies from $2.9 \times 10^{-6}$ to $2.7 \times 10^{-6}$ s$^{-1}$ between 0° and 30° heliolatitude. The value of $\Lambda$ is taken to be the heliographic latitude of the foot point of the nominal Sun-Earth field line.

Equation (7) provides a first order approximation of the path length. Using a zero proton pitch angle (because of the rapid focussing of a divergent field) the path length for GLE 69 is estimated at 1.1 AU, based on a mean solar wind speed $V_{SW}$ of ~600 km s$^{-1}$. For this event no significant flux above 7.6 GV (6.7 GeV) was detected by neutron monitors at sea level and the injection time at the Sun was calculated based on this energy (i.e., $\beta = 0.99$). We calculate the travel time to be 540±40 seconds. To estimate the injection time at the Sun we simply subtract this travel time from the onset time of GLE 69 (06:48 UT ±30 seconds, 1 minute data), to give 06:39 ST ±50 seconds. These results include a 10% uncertainty in the mean solar wind speed.

For GLE 60, the path length along the IMF is estimated at 1.2 AU, based on a mean solar wind speed $V_{SW}$ of ~400 km s$^{-1}$. This result is in agreement with Bieber et al. (2004). In contrast, Sáiz et al. (2005a), using the "inverse velocity method", calculated a path length of 1.7 AU. However, they did not have confidence in this result and considered the result of Bieber et al. (2004) (which uses a detailed interplanetary transport model) to be more accurate. No significant flux above 6.3 GV (5.4 GeV) was detected by neutron monitors at sea level. We calculate a travel time at this energy and path length of 1.2 AU to be 590±85 seconds. Based on the neutron monitor onset time of 13:55 UT ±30 seconds, we estimate the injection time at the Sun to be 13:45 ST ±90 seconds.

Bieber et al. (2004) and Sáiz et al. (2005a, 2005b) derived relativistic proton injection times for GLEs 60 and 69 within 3 minutes of our estimates.

For GLE 59, the path length along the IMF is estimated at 1.1 AU, based on a mean solar wind speed $V_{SW}$ of ~600 km s$^{-1}$. No significant flux above 3 GV (2.2 GeV) was detected by sea-level neutron monitors. We calculate a proton travel time at this energy (i.e., $\beta = 0.95$) as 565±40 seconds. Based on a neutron monitor onset time of 10:32 UT ±30 seconds, we estimate the injection time at the Sun to be 10:23 ST ±50 seconds.

## 5. DISCUSSION

### 5.1. *State of the Interplanetary Medium: GLE 69*

A series of solar eruptions occurred during a period of intense solar activity extending from 14 to 20 January 2005. ACE measurements of the IMF strength for the period indicate the passage of shocks and associated magnetic structures suggesting that the IMF had indeed experienced substantial disturbances. However, by the time the GLE commenced (06:48 UT) the intensity of the IMF had recovered somewhat with ACE measuring an average magnetic field strength of ~5 nT.

Observations of solar wind speeds by WIND and ACE were affected by the intense particle emission. However, the CELIAS proton detector on board the SOHO spacecraft measured a solar wind speed ranging from 600 to 800 km s$^{-1}$ (hourly averages) in the 36 hours between the GLE onset and the arrival of the shock at Earth.

Particle pitch angle distributions may provide information about the interplanetary medium through which the particles have travelled. The degree of anisotropy contains important information on scattering effects in the interplanetary medium. For GLE 69 the particle arrival at 06:55 UT (peak) was strongly anisotropic, indicating focused transport conditions. After ~07:00 UT the field-aligned component of the pitch angle distribution began to broaden and, in addition, local scattering began to increase.

Pitch angle distributions from 07:20 to 08:00 UT show a slight indication of an enhancement near 180° (Figs. 5b and 5c). We therefore examined the possibility of bi-directional flow, incorporating a modification of the pitch angle distribution function of equation (1) as follows:

$$G'(\alpha) = G_1(\alpha) + C \times G_2(\alpha') \tag{8}$$

where $G_1$ and $G_2$ are of the same form as in equation (1) with independent parameters $A_1$, $B_1$, $A_2$ and $B_2$; $\alpha' = \pi - \alpha$; and $C$ is the ratio of reverse-to-forward flux ranging from 0 and 1. (see Cramp et al. 1997*a* and 1997*b* for a review of the method).

Bi-directional flow could result from: (1) enhanced turbulence in a disturbed interplanetary medium beyond Earth's orbit which results in the back-scattering of particles; or (2) the result of particles arriving from the Sun along two different paths of a closed interplanetary magnetic loop. Magnetic mirroring, as evidenced by bi-directional flow of relativistic particles following intense solar activity, has been previously reported by Cramp et al. (1997a) and Bieber et al. (2002) and bi-directional flow in a closed interplanetary magnetic loop has been reported by Ruffolo et al. (2006).

Our modelling shows that by 08:00 UT (Fig. 9) there is evidence of an excess of reverse-propagating particles. Table (10) shows the improved fits to observations when bi-directional flow parameters are included in the modelling. This enhancement could be attributed to turbulence in the disturbed interplanetary medium beyond Earth's orbit resulting from previous solar activity described above. Alternatively, the enhancement could be attributed to particles arriving from the Sun along two different paths in a closed interplanetary magnetic loop configuration. We are unaware of any evidence in the literature to support the existence of a closed interplanetary magnetic loop configuration for this event. This suggests that back scattering from a reflecting boundary beyond Earth is a more likely cause for the bi-directional flow.

5.2. *Source Mechanism: GLE 69*

The location of the intense emissions (N14° W61°) for the 2005 January 20 solar eruption must have been close to the nominal Sun-Earth connecting magnetic field line. The extremely rapid rise (~5 minutes to peak intensity) in neutron monitor count rates is evidence for excellent connectivity. The strong anisotropy near GLE onset indicates that relativistic particles travelled along the IMF essentially scatter-free. Bieber et al. (2005) noted that over a 6-minute interval the neutron monitor count rate (in 1-minute data) at South Pole, McMurdo and Terre Adélie increased by factors of 56, 30 and 46 respectively. For the same interval other stations observed increases of only a factor of 3. This suggests that relativistic protons arriving at 1 AU were initially confined to a narrow beam. These characteristics provide a rare opportunity to directly observe the relativistic proton acceleration source from Earth during the initial phase of the event and allow us to test theoretical particle acceleration models. In a preliminary study, Sáiz et al. (2005b) have suggested that such intense particle emission may have interacted with Alfvén waves, transferring energy to the medium. The effect of such a mechanism on the particle spectrum is unclear.

Of the spectra employed, our results show that neutron monitor observations from the peak (06:55 UT) to the decline (07:30 UT) phases of the event are best fitted by the stochastic model. Mechanisms for relativistic particle acceleration via stochastic processes include magnetic reconnection in solar flares and coronal neutral current sheet reconnection behind an erupting CME (e.g., Lin and Forbes 2000; Klein et al. 2001; and Lin et al. 2005).

This event was also associated with a very fast CME with speed estimates ranging from 2500 to 3242 km/s (Gopalswamy et al. 2005, Simnett 2006). The variable parameters in the Ellison and Ramaty equation are the shock compression ratio r and the e-folding energy $E_0$ (MeV). Ellison and Ramaty (1985) showed that the proton spectrum between 1 MeV and 10 GeV for various SPE and GLE events can be fitted with compression ratios ranging from ~1.6 to 3.0. At 06:55 UT when the 20 January CME was low in the corona, the Ellison and Ramaty fit gave a compression ratio of 1.4 increasing to 1.6 at 07:30. Whilst these values of the compression ratio are smaller than those reported in Papers 1 and 2 for the 2000 July 14 and 2001 April 15 CMEs (1.93 and 2.60 respectively), they do fall in the lower range of values found by Ellison and Ramaty for relativistic particle acceleration. However, the values of *e*-folding energy (Table 4) are generally larger than we have seen in other events, indicating that spectra with rapid roll-off below 10 GeV cannot fit the observed data.

Recent theoretical investigations by Lee (2005) and Giacalone (2005) have shown that the high-energy part of the spectrum will be dominated by particles produced when the shock is quasi-perpendicular (Tylka and Lee 2006). More importantly, Lee (2005) and Giacalone (2005) show that the roll-over at higher energies is not exponential, as assumed by the Ellison & Ramaty (1985). In particular, for this event, modified power law spectra show very little roll-over and it is possible that the shock models of Lee (2005) and Giacalone (2005) could potentially provide better fits to the observations than are achieved by the Ellison & Ramaty spectral form. Therefore, one must be cautious not to rule out particle acceleration via a CME–driven shock for this event. However, even if we consider the limitations of the Ellison & Ramaty spectral form, the stochastic model appears to better represent the neutron monitor observations.

Kuznetsov et al. (2005) and Simnett (2006) considered the relative timing of various electromagnetic signals (e.g., microwave to gamma ray emissions) for this event as well as the onset time of the GLE. They suggest that relativistic proton acceleration occurred during the main flare phase and it was this phase that was responsible for the leading spike in neutron monitor observations. Simnett (2006) concluded that at least during the initial stages of this solar event that the 20 January CME was not responsible for relativistic proton acceleration. However, relativistic proton acceleration continued for several hours after the main flare phase ended, as is evident by global neutron monitor responses. This indicates that there was an additional source of relativistic particles, possibly the 20 January CME.

5.3. *Source Mechanisms: GLEs 59 and 60*

In light of our revised modelling approach we have shown that the model derived parameters for GLEs 59 and 60 are similar to our previous investigations (see Papers 1 and 2). Therefore, our conclusions in regard to relativistic proton production for these two events remain unchanged. For GLE 59 our modelling indicates that more than one acceleration mechanism was present, consistent with both shock and stochastic acceleration processes dominating at different times of the event. For GLE 60 we find that Ellison and Ramaty spectra better represent the neutron monitor observations compared to stochastic acceleration spectra.

*5.4. Comparisons with GOES observations: GLEs 59, 60 and 69*

Spacecraft data have been corrected for arrival-time velocity dispersion by simply shifting the GOES proton time-lines backwards by the difference in transit time estimates (see section 4.5). Figure 10 shows that particle intensities at spacecraft energies (ranging from ~30 to ~100 MeV) are considerably lower than predicted from fitted spectra for all three events. Using GLE 59 as an example, a 30-MeV particle injected into a smooth interplanetary medium at 10:23 ST would require a travel time of ~37 minutes, arriving at 1 AU at ~11:00 UT. Figure 10a shows that at 11:10 UT 30-MeV proton intensities remain significantly lower than expected values. This suggests that the propagation of low-energy protons en-route to Earth could have been affected by processes which result in the trapping of particles at the source and/or other interplanetary transport effects outlined in section 4. By excluding spacecraft data we have minimised the impact of such effects on our analyses.

6. CONCLUSION

We have modelled the arrival of relativistic protons at 1 AU for the 2005 January 20 solar event. The neutron monitor intensity/time profiles show a rapid rise to maximum in approximately 5 minutes, indicating that relativistic protons had excellent access to Sun-Earth connected magnetic field lines. We find that the event was marked by a highly anisotropic onset followed by a fairly rapid decrease in anisotropy. We attribute part of the underlying isotropic component in pitch angle distributions to bi-directional flow and propose that back-scattering from a reflecting boundary beyond Earth is a likely cause.

We employed shock and stochastic acceleration spectral forms in our fits to neutron monitor observations. We find that from the peak (06:55 UT) to the decline (07:10 UT) phases of GLE 69, neutron monitor observations are better fitted by the Gallegos-Cruz & Perez-Peraza (1995) stochastic acceleration model.

We have also re-examined GLEs 59 and 60. For GLE 59 our modelling indicates that more than one acceleration mechanism was present, consistent with both shock and stochastic acceleration processes dominating at different times of the event. For GLE 60 we reaffirm our earlier findings that Ellison and Ramaty spectra better represent the neutron monitor observations compared to stochastic acceleration spectra. Furthermore, we provide evidence that for each of these events low-energy proton intensities were depressed, which could be due to transport processes.

A future study should compare the recent Lee (2005) and Giacalone (2005) models of diffusive shock acceleration with the widely used Ellison and Ramaty (1985) spectral form to determine whether or not they produce fits comparable to the Gallegos-Cruz & Perez-Peraza (1995) stochastic acceleration model.

ACKNOWLEGMENTS

We are grateful to the referees for their extensive and valuable comments. We thank our colleagues at IZMIRAN (Russia) and The Polar Geophysical Institute (Russia) for contributing neutron monitor data. D. J. Bombardieri acknowledges receipt of an Australian Postgraduate Award and Australian Antarctic Science Scholarship. Neutron monitors of the Bartol Research Institute are supported by NSF ATM-0527878. GOES 11 data were obtained from Space Physics Interactive Data Resource http://spidr.ngdc.noaa.gov/.

FIGURE CAPTIONS

FIG. 1. ------ Solar cosmic ray intensity/time profiles (corrected to sea-level pressure) for 2005 January 20 as recorded by Terre Adélie, McMurdo, South Pole, SANAE, Mawson and Jungfraujoch neutron monitors. The percentage increase for Jungfraujoch is shown on a different scale. The impulsive nature of the neutron monitor intensity/time profiles for Terre Adélie, McMurdo and South Pole and the rapid rise (5 minutes) to maximum is typical of well-connected events.

FIG. 2. ------ Viewing directions of neutron monitors in geographic coordinates at 06:55 UT (peak) on 20 January 2005. Geomagnetic conditions were quiet (Kp = 4; DST = -58). Lines for each station represent the vertical viewing direction at different rigidities. Numeral 5 represents the vertical viewing direction at maximum rigidity (~5 GV), while numeral 1 represents the vertical viewing direction at the atmospheric cutoff (~1 GV). The small filled black circles represent 1 GV increments. The large filled black circle represents the IMF direction as measured by ACE at 06:55 UT. Station abbreviations are: APT = Apatity, Russia; BBG =

Barensburg, Russia; CPS = Cape Schmidt, Russia; IVK = Inuvik, Canada; MAW = Mawson, Antarctica; NAI = Nain, Canada; MCM = McMurdo, Antarctica; SAN = SANAE, Antarctica; SPO = South Pole, Antarctica; TER = Terre Adélie, Antarctica; THU = Thule, Greenland; TXB = Tixie Bay, Russia.

FIG. 3. ------ Observed (solid circles) and modelled (lines) responses to the 20 January 2005 GLE. The percentage increase for Mawson, Nain, Apatity and SANAE are shown on a different scale to South Pole and McMurdo.

FIG. 4. ------ GSE longitude (top) and GSE latitude (bottom) of the apparent arrival directions (solid circles) plotted with the negative magnetic field direction (1 hour centred moving averages) as measured by the ACE spacecraft (open circles).

FIG. 5. ------ Derived pitch angle distributions for the GLE of 2005 January 20: (a) 06:50 UT, 06:55 UT, 07:00 UT; (b) 07:10 UT, 07:20 UT, 07:30 UT; (c) 07:40 UT, 07:50 UT 08:00 UT.

FIG. 6. ------ Derived rigidity spectra using a modified power law (see text) for the GLE of 2005 January 20: 06:50 UT, 06:55 UT, 07:00 UT, 07:05 UT, 07:10 UT and 07:15 UT.

FIG. 7. ------ Energy spectral fits to ground-based neutron monitor observations (ranging from ~400 MeV to 10 GeV) for intervals 07:00 UT, 07:10 UT 07:20 UT 07:30 UT. Neutron monitor observations (black dots) are shown with corresponding 1-sigma error bars. Fitted curves are of the Ellison & Ramaty (1985) spectral form (light grey line) and the Gallegos-Cruz & Perez-Peraza (1995) stochastic acceleration spectral form (black line).

FIG. 8. ------ Residuals from the non-linear least squares fit plotted against kinetic energy for GLE 69 at 07:10 UT. Stochastic acceleration model (black line) and the Ellison & Ramaty (1985) spectral form (light grey line).

FIG. 9. ------ Derived pitch angle distributions incorporating bi-directional flow parameters for 06:55 UT, 07:10 UT, 07:40 UT, and 08:00 UT. Note enhancement at pitch angles near 180° for the 07:40 UT and 08:00 UT time intervals, which is indicative of bi-directional flow.

FIG. 10. ------ Energy spectral fits to ground based neutron monitor observations: (a) 11:10 UT, (b) 14:45 UT and (c) 07:10 UT decline phases for GLEs 59, 60 and 69 respectively. Five-minute proton data (open triangles) from GOES 8, 10 and 11 energetic particle sensors (EPS) have been corrected for arrival time velocity dispersion; energy range is ~30 to ~330 MeV. Neutron monitor derived data for (a) (small black circles) ranges from ~400 MeV to 5 GeV. Neutron monitor derived data for (b) (small black circles) ranges from ~400 MeV to 10 GeV. Fitted curves are of the Ellison & Ramaty (1985) spectral form (light grey line) and the Gallegos-Cruz & Perez-Peraza (1995) stochastic acceleration spectral forms (black line). Note that intensity of low-energy protons (30 to 100 MeV) for (a) (b) and (c) are significantly less than theoretical estimates.

TABLE 1

NEUTRON MONITOR AND GEOMAGNETIC CUTOFF RIGIDITIES

| Station | Lat. (deg.) | Lon. (deg.) | Alt. (m) | $P_c^a$ (GV) | Onset[b] UT[b] |
|---|---|---|---|---|---|
| Alma Ata……… | 43.25 | 76.92 | 3340 | 6.69 | - |
| Apatity………… | 67.55 | 33.33 | 177 | 0.61 | 06:51 |
| Athens………… | 37.97 | 23.72 | 40 | 8.72 | - |
| Baksan………… | 43.28 | 42.69 | 1700 | 5.70 | - |
| Barensburg…….. | 78.06 | 14.22 | 0 | 0.07 | 06:57 |
| Bern…………… | 46.55 | 7.98 | 570 | 4.42 | - |
| Calgary………… | 51.08 | 245.87 | 1128 | 1.09 | - |
| Cape Schmidt….. | 68.92 | 180.53 | 0 | 0.45 | 06:57 |
| Climax………… | 39.37 | 253.82 | 3400 | 3.03 | 06:50 |
| Durham………… | 43.10 | 289.17 | 0 | 1.58 | - |
| Erevan………… | 40.50 | 44.17 | 3200 | 7.58 | 06:57 |
| Fort Smith……… | 60.00 | 258.10 | 0 | 0.30 | 06:52 |
| Hermanus………. | -34.42 | 19.22 | 26 | 4.90 | - |
| Inuvik………….. | 68.35 | 226.28 | 21 | 0.18 | 06:56 |
| Irkustk………….. | 52.28 | 104.02 | 435 | 3.66 | 06:51 |
| Jungfraujoch…… | 46.55 | 7.98 | 3475 | 4.48 | 06:52 |
| Kerguelen Island.. | -49.35 | 70.25 | 33 | 1.19 | 06:51 |
| Kiel……………. | 54.33 | 10.13 | 54 | 2.29 | 06:52 |
| Kingston……….. | -42.99 | 147.29 | 65 | 1.88 | 06:53 |
| LARC………….. | -62.20 | 301.04 | 40 | 2.21 | 06:53 |
| Lomnický Ŝtít….. | 49.20 | 20.22 | 2634 | 4.00 | 06:53 |
| Magadan……….. | 60.12 | 151.02 | 220 | 2.10 | 06:54 |
| Mawson………... | -67.60 | 62.88 | 30 | 0.22 | 06:51 |
| McMurdo………. | -77.85 | 166.72 | 48 | 0.01 | 06:49 |
| Mexico City……. | 19.33 | 260.80 | 2274 | 8.61 | - |
| Moscow………... | 55.47 | 37.32 | 200 | 2.46 | 06:53 |
| Mt. Washington... | 44.30 | 288.70 | 1909 | 1.46 | - |
| Nain……………. | 56.55 | 298.32 | 0 | 0.45 | 06:51 |
| Newark………… | 39.68 | 284.25 | 50 | 1.97 | 06:50 |
| Norlisk…………. | 69.26 | 88.05 | 0 | 0.63 | 06:52 |
| Novosibirsk……. | 54.80 | 83.00 | 163 | 2.91 | 06:56 |
| Oulu……………. | 65.05 | 25.47 | 15 | 0.81 | 06:51 |
| Potchefstrom…… | -26.68 | 27.10 | 1351 | 7.30 | - |
| Rome…………... | 41.86 | 12.47 | 60 | 6.32 | - |
| SANAE………… | -71.67 | 357.15 | 856 | 1.06 | 06:49 |
| South Pole……… | -90.00 | 0.00 | 2820 | 0.10 | 06:48 |
| Terre Adélie........ | -66.67 | 140.02 | 45 | 0.01 | 06:49 |
| Thule…………… | 76.50 | 291.30 | 260 | 0.00 | 06:55 |
| Tsumeb………… | -19.20 | 17.58 | 1240 | 9.29 | - |
| Tixie Bay………. | 71.58 | 128.92 | 0 | 0.53 | 06:54 |
| Yakutsk………… | 62.03 | 129.73 | 105 | 1.70 | 06:52 |

[a] Nominal vertical geomagnetic cut-off rigidities represent the minium rigidities below which particles do not have access to a particular site on the Earth's surface. The cut-off at the geomagnetic equator is ~17 GV, decreasing to zero at the geomagnetic poles.
[b] GLE onset times (1-minute data). Missing intervals (denoted by dash) indicate either 1-minute data not available or station did not measure a significant increase.

TABLE 2

PITCH ANGLE DISTRIBUTION PARAMETERS
20 JANUARY 2005

| Time [a] (UT) | A [b] | error | B [c] | error |
|---|---|---|---|---|
| 06:50…… | 0.308 | ±0.055 | 0.000 | ±0.000 [d] |
| 06:55…… | 0.319 | ±0.057 | 0.100 | ±0.010 |
| 07:00…… | 0.287 | ±0.052 | 0.504 | ±0.050 |
| 07:10…… | 0.354 | ±0.064 | 1.082 | ±0.108 |
| 07:20…… | 0.146 | ±0.026 | 2.145 | ±0.215 |
| 07:30…… | 0.085 | ±0.015 | 2.321 | ±0.232 |
| 07:40…… | 0.712 | ±0.130 | 2.820 | ±0.282 |
| 07:50…… | 0.235 | ±0.042 | 2.334 | ±0.233 |
| 08:00…… | 0.408 | ±0.073 | 4.128 | ±0.413 |

[a] Time refers to the start of a five-minute interval.

[b] Parameter A has most effect on the width of the anisotropy.

[c] Parameter B has most effect on the relative flux in the reverse direction.

[d] The error is zero quoted to the number of significant figures

TABLE 3

SPECTRAL PARAMETERS
20 JANUARY 2005

| Time [a] (UT) | $J_{\parallel}$ [b] | $\gamma$ [c] | $\delta\gamma$ [d] |
|---|---|---|---|
| 06:50……. | 2244.2 | -5.5 ±0.8 | 0.045±0.009 |
| 06:55……. | 20451.4 | -9.2 ±0.6 | 0.072±0.014 |
| 07:00……. | 4694.3 | -7.7±0.5 | 0.000±0.000 [e] |
| 07:05……. | 2531.8 | -7.4±0.5 | 0.000±0.000 [e] |
| 07:10……. | 1505.1 | -7.3 ±0.5 | 0.000±0.000 [e] |
| 07:15……. | 1508.4 | -7.3 ±0.5 | 0.010±0.002 |

[a] Time (UT) refers to the start of a five-minute interval.
[b] $J_{\parallel}$ is the particle flux ((cm$^2$ s sr GV)$^{-1}$) at 1 GV summed over the forward steradian.
[c] Spectral slope ($\gamma$) with corresponding errors.
[d] Modified power law exponent modifier ($\delta\gamma$) with corresponding errors.
[e] The error is zero quoted to the number of significant figures

TABLE 4

VARIABLE MODEL PARAMETERS
ELLISON AND RAMATY SPECTRAL FORM
20 JANUARY 2005

| Time[a] (UT) | r[b] | $E_0$[c] (MeV) | WSS[d] |
|---|---|---|---|
| 06:55…….. | 1.417 ±0.002 | 5100 ±485 | 46.0 |
| 07:00…….. | 1.491 ±0.003 | 9372 ±213 | 35.8 |
| 07:05…….. | 1.527 ±0.003 | 8298 ±995 | 39.2 |
| 07:10…….. | 1.559 ±0.005 | 6383 ±685 | 53.2 |
| 07:15…….. | 1.583 ±0.006 | 4932 ±470 | 70.0 |
| 07:20…….. | 1.602 ±0.006 | 4432 ±402 | 78.4 |
| 07:25…….. | 1.601 ±0.007 | 4046 ±68 | 95.0 |
| 07:30…….. | 1.609 ±0.008 | 3525 ±309 | 116.2 |

[a] Time refers to the start of a five-minute interval.
[b] Shock compression ratio.
[c] e-folding energy.
[d] Weighted sum of squares.

TABLE 5

VARIABLE MODEL PARAMETERS
STOCHASTIC ACCELERATION MODEL
20 JANUARY 2005

| Time[a] (UT) | N[b] | $\alpha$[c] ($s^{-1}$) | WSS[d] |
|---|---|---|---|
| 06:55……… | 1968 ±98 | 0.0146 ±0.0001 | 3.7 |
| 07:00……… | 72 ±4.0 | 0.0215 ±0.0003 | 6.4 |
| 07:05……… | 27 ±5.8 | 0.0236 ±0.0003 | 6.0 |
| 07:10……… | 14 ±1.0 | 0.0242 ±0.0003 | 5.9 |
| 07:15……… | 11 ±0.6 | 0.0237 ±0.0003 | 5.4 |
| 07:20……… | 8 ±0.4 | 0.0238 ±0.0002 | 4.9 |
| 07:25……… | 10 ±0.6 | 0.0280 ±0.0002 | 6.1 |
| 07:30……… | 11 ±1.0 | 0.0218 ±0.0002 | 6.4 |

[a] Time refers to the start of a five-minute interval.
[b] Normalization factor.
[c] Acceleration efficiency.
[d] Weighted sum of squares.

TABLE 6

VARIABLE MODEL PARAMETERS
ELLISON AND RAMATY SPECTRAL FORM
14 JULY 2000

| Time [a] (UT) | r [b] | $E_0$ [c] (MeV) | WSS [d] |
|---|---|---|---|
| 10:45…….. | 1.93 ±0.02 | 1869 ±6 | 88 |
| 10:55……. | 1.78 ±0.01 | 1723 ±55 | 154 |
| 11:00……. | 1.76 ±0.02 | 1527 ±72 | 209 |
| 11:10…….. | 1.72 ±0.02 | 1442 ±68 | 234 |
| 11:40……. | 1.75 ±0.03 | 1211±46 | 1047 |

[a] Time refers to the start of a five-minute interval.
[b] Shock compression ratio.
[c] e-folding energy
[d] Weighted sum of squares

TABLE 7

VARIABLE MODEL PARAMETERS
STOCHASTIC ACCELERATION MODEL
14 JULY 2000

| Time [a] (UT) | N [b] | $\alpha$ [c] ($s^{-1}$) | WSS [d] |
|---|---|---|---|
| 10:45………. | 0.30 ±0.09 | 0.0300±0.0010 | 671 |
| 10:55…….. | 1.45 ±0.03 | 0.0234±0.0001 | 2 |
| 11:00…….. | 2.10 ±0.01 | 0.0216±0.0002 | 11 |
| 11:10…….. | 4.30 ±0.30 | 0.0195±0.0002 | 12 |
| 11:40…….. | 9.08 ±0.80 | 0.0164±0.0002 | 35 |

[a] Time refers to the start of a five-minute interval.
[b] Normalization factor.
[c] Acceleration efficiency.
[d] Weighted sum of squares.

TABLE 8

VARIABLE MODEL PARAMETERS
ELLISON AND RAMATY SPECTRAL FORM
15 APRIL 2001

| Time[a] (UT) | r [b] | $E_0$ [c] (MeV) | WSS [d] |
|---|---|---|---|
| 14:10…….. | 2.60 ±0.02 | 2667 ±107 | 835 |
| 14:20…….. | 2.05 ±0.02 | 3045 ±57 | 51 |
| 14:30…….. | 1.86 ±0.01 | 4104 ±30 | 10 |
| 14:45…….. | 1.83 ±0.01 | 3559 ±29 | 31 |

[a] Time refers to the start of a five-minute interval.
[b] Shock compression ratio.
[c] e-folding energy.
[d] Weighted sum of squares

TABLE 9

VARIABLE MODEL PARAMETERS
STOCHASTIC ACCELERATION MODEL
15 APRIL 2001

| Time[a] (UT) | N [b] | $\alpha$ [c] ($s^{-1}$) | WSS [d] |
|---|---|---|---|
| 14:10…….. | 0.01 ±0.01 | 0.045 ±0.008 | 2976 |
| 14:20…….. | 0.05 ±0.01 | 0.041 ±0.002 | 251 |
| 14:30…….. | 0.20 ±0.03 | 0.036 ±0.001 | 84 |
| 14:45…….. | 0.30 ±0.01 | 0.035 ±0.001 | 54 |

[a] Time refers to the start of a five-minute interval.
[b] Normalization factor.
[c] Acceleration efficiency.
[d] Weighted sum of squares.

TABLE 10

PITCH ANGLE DISTRIBUTION
EFFECT OF BI-DIRECTIONAL FLOW

| Time[a] (UT) | WSS [b] | WSS [c] |
|---|---|---|
| 07:40…….. | 2500 | 2440 |
| 07:50…….. | 1720 | 1500 |
| 08:00…….. | 1120 | 910 |

[a] Time refers to the start of a five-minute interval.
[b] Weighted sum of squares: Standard pitch angle distribution
[c] Weighted sum of squares: Bi-directional flow

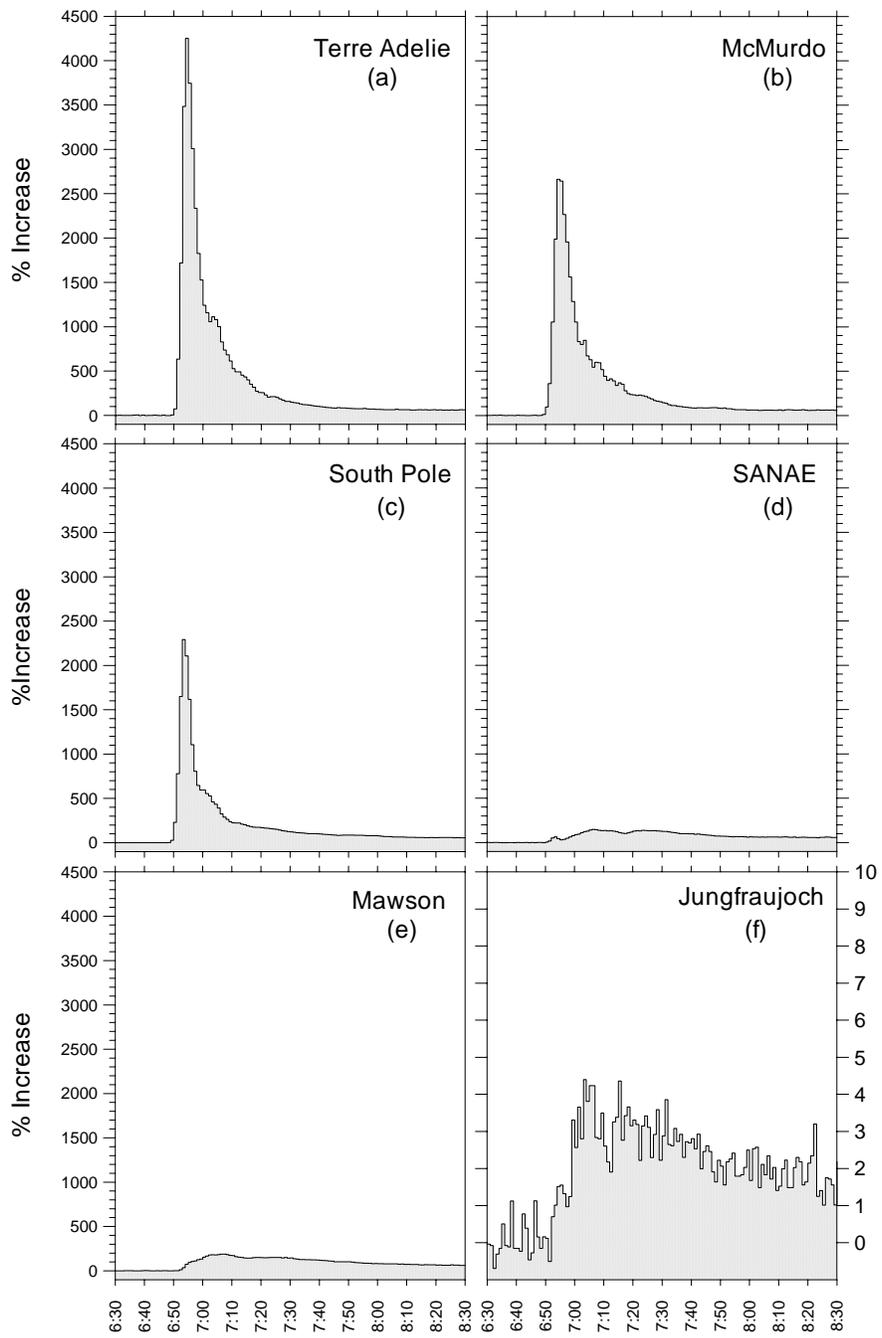

FIGURE 1

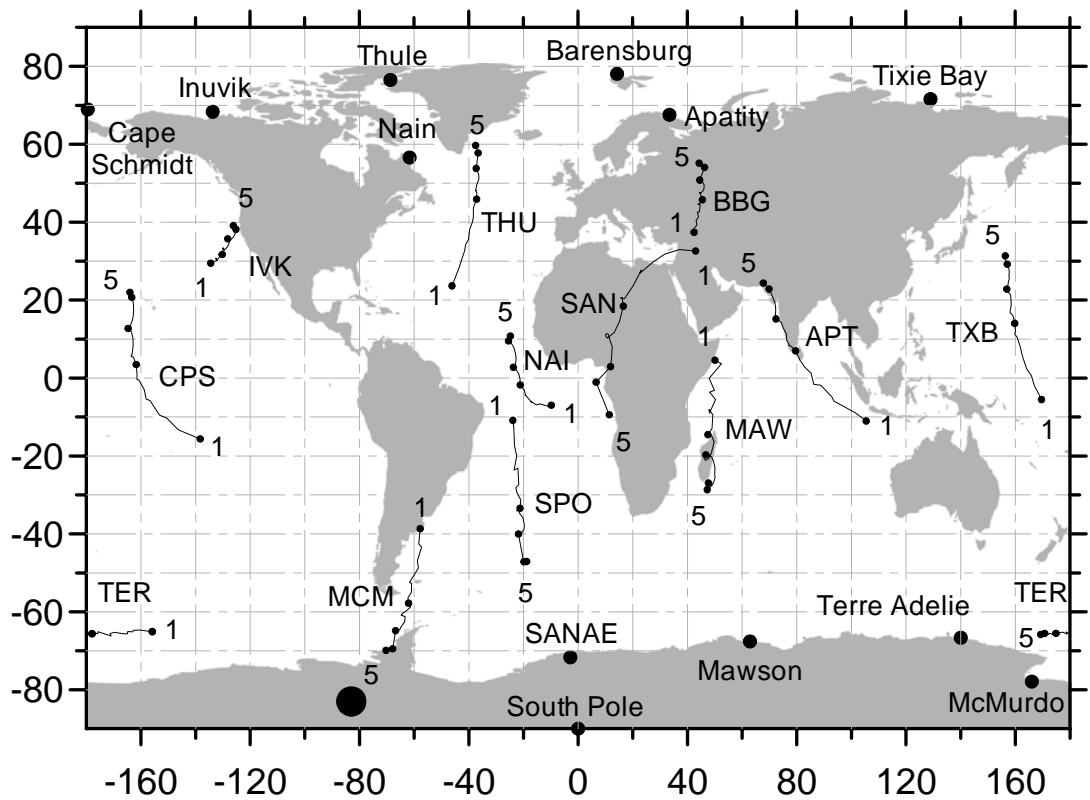

FIGURE 2

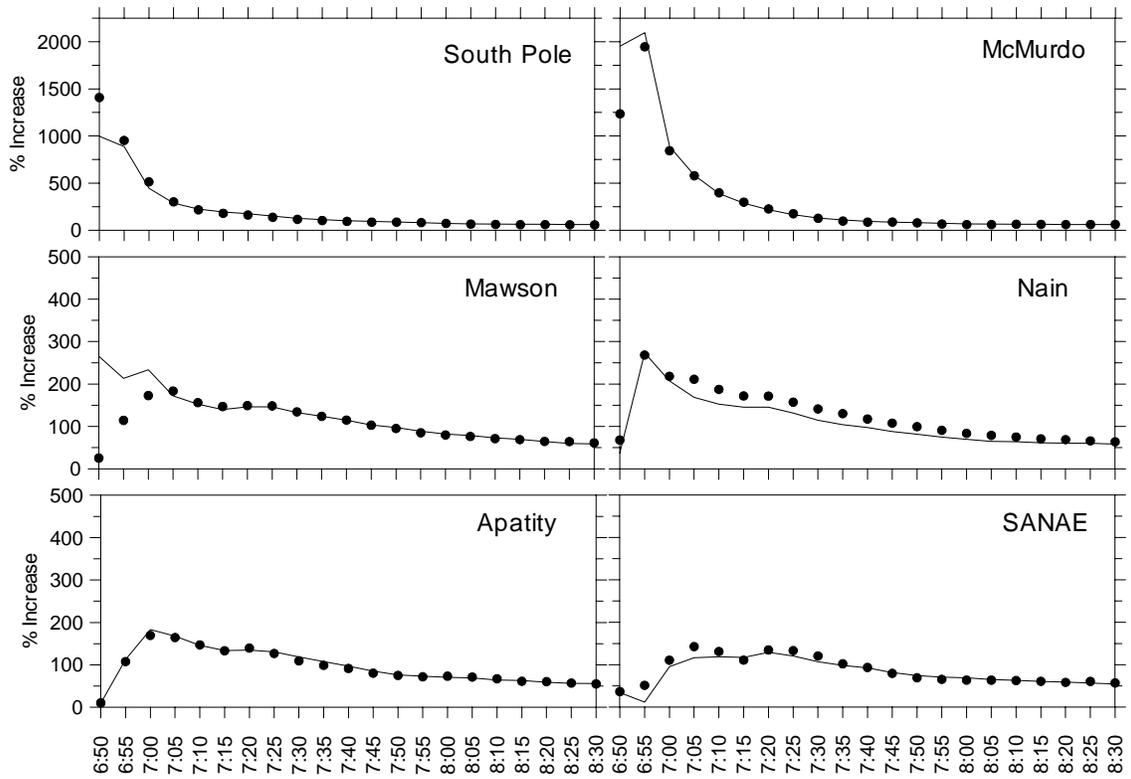

FIGURE 3

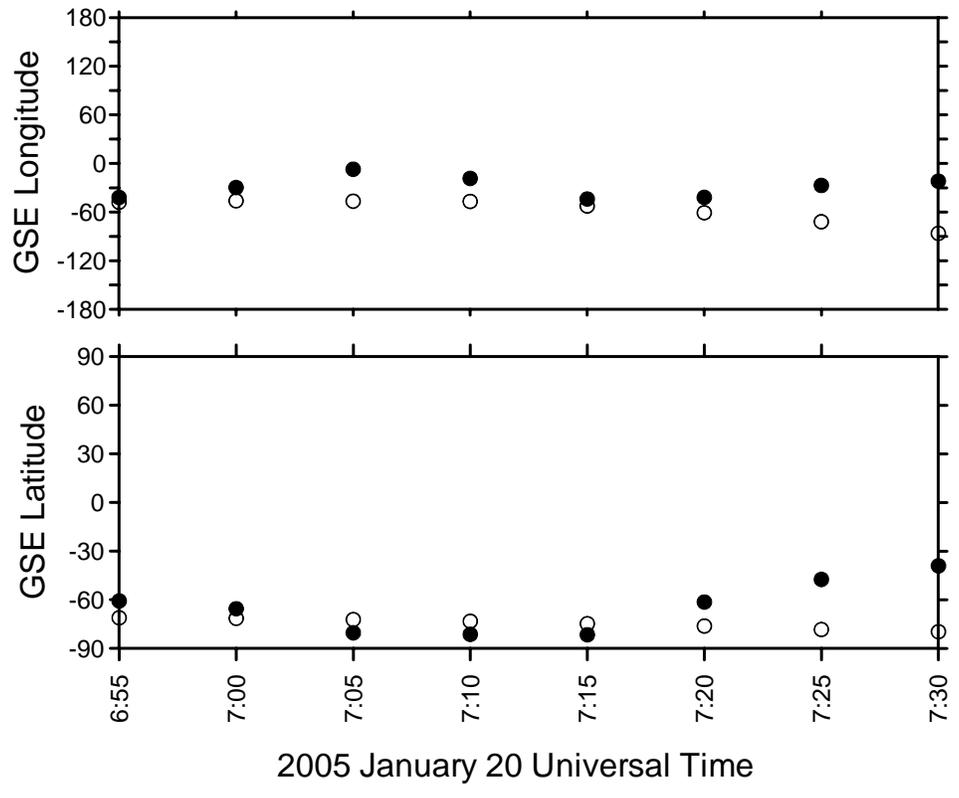

FIGURE 4

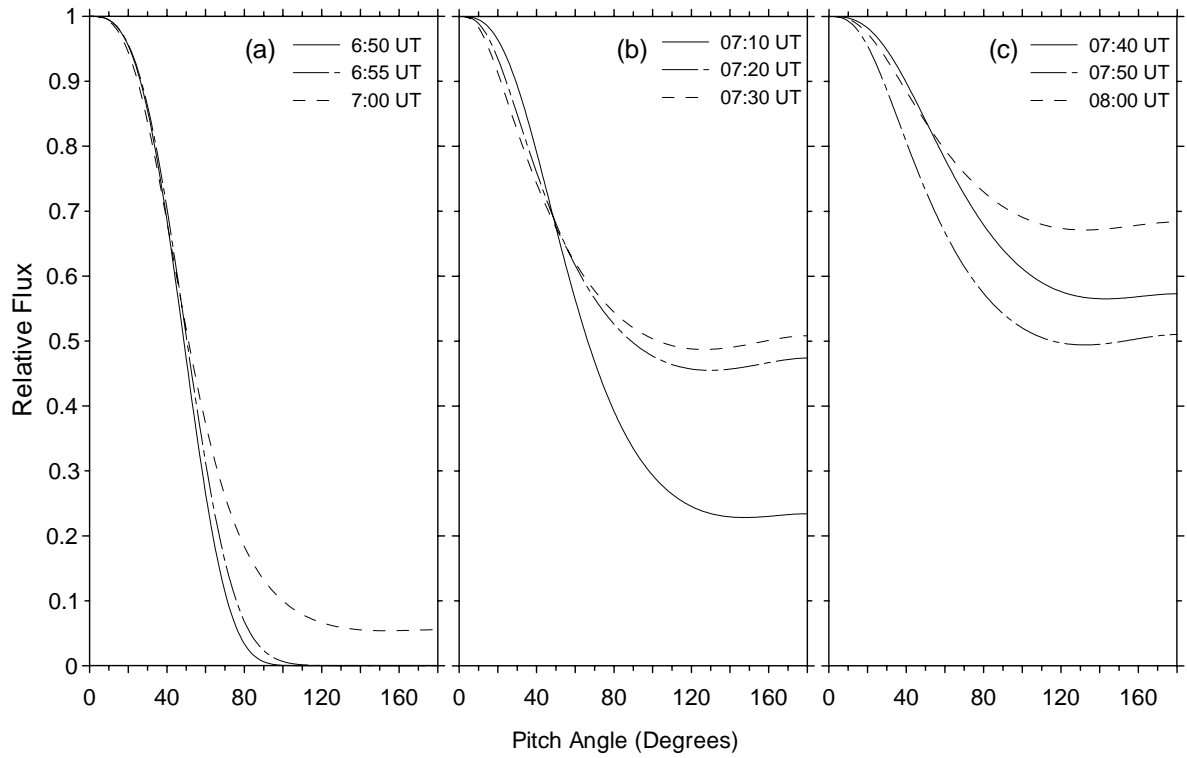

FIGURE 5

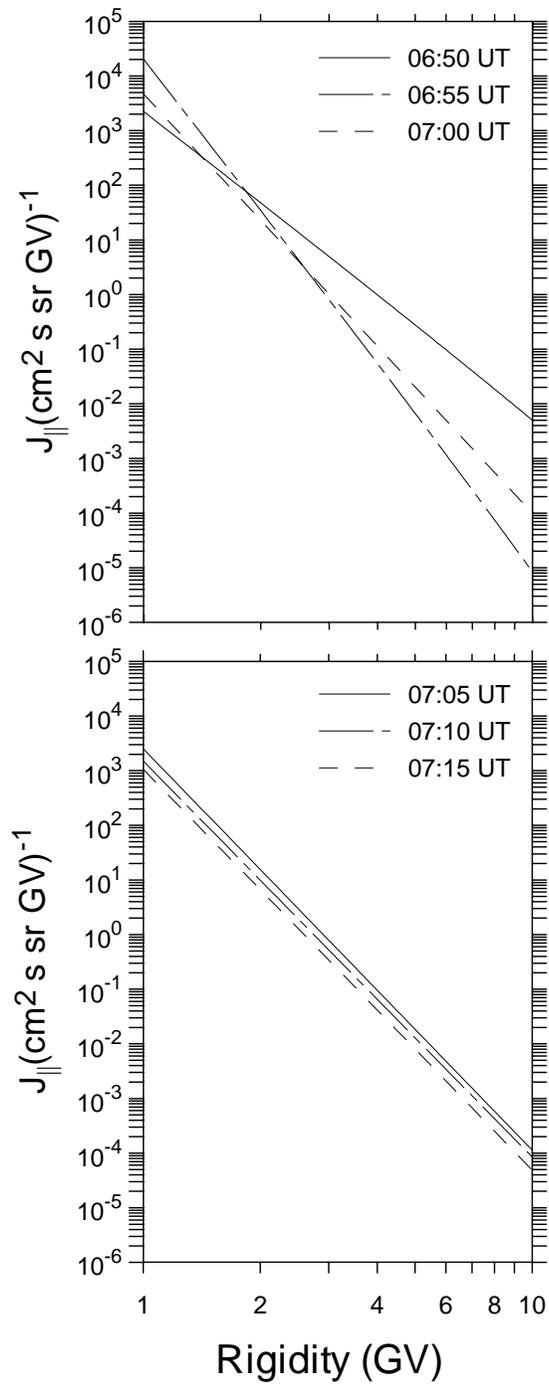

FIGURE 6

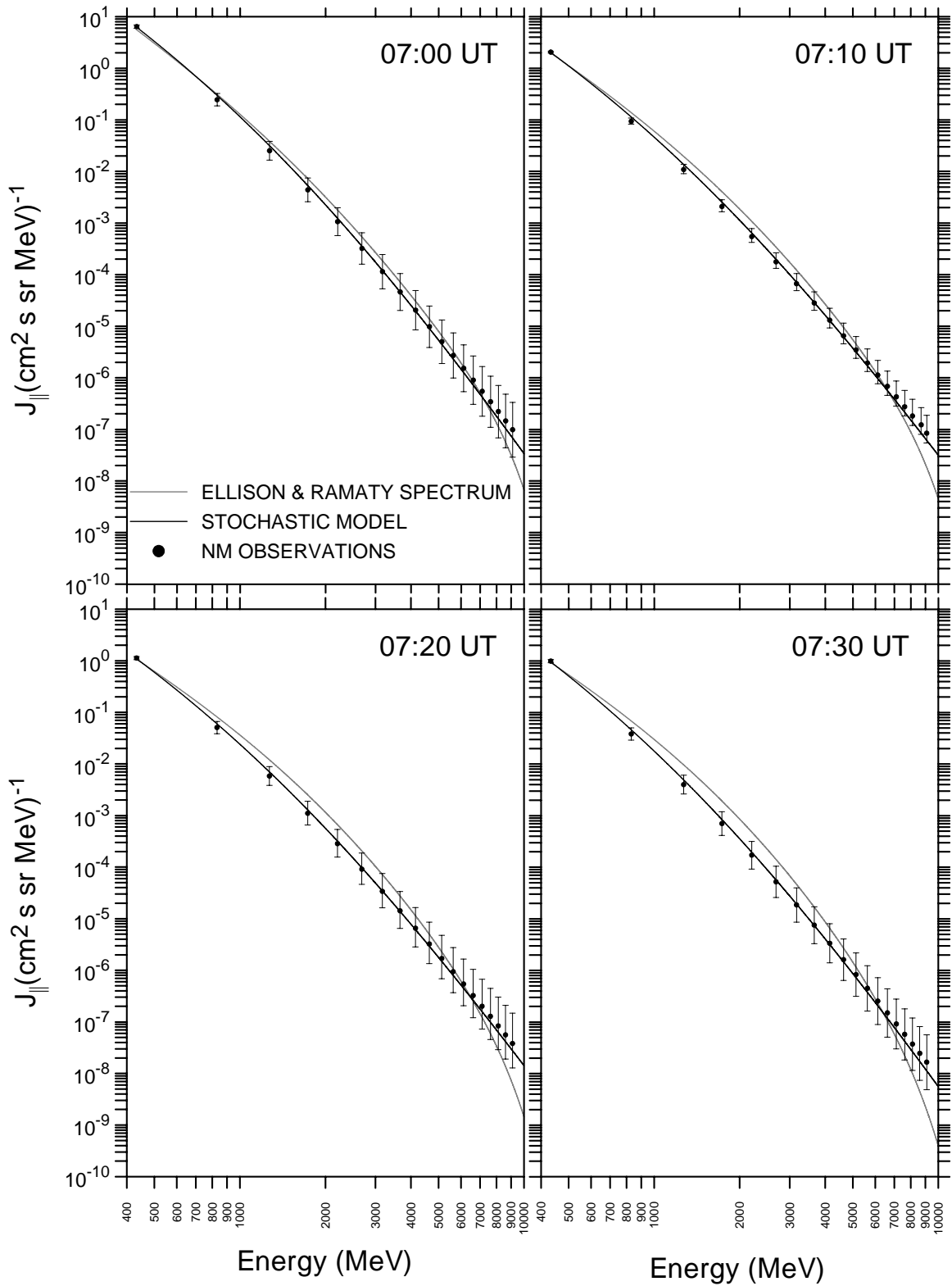

FIGURE 7

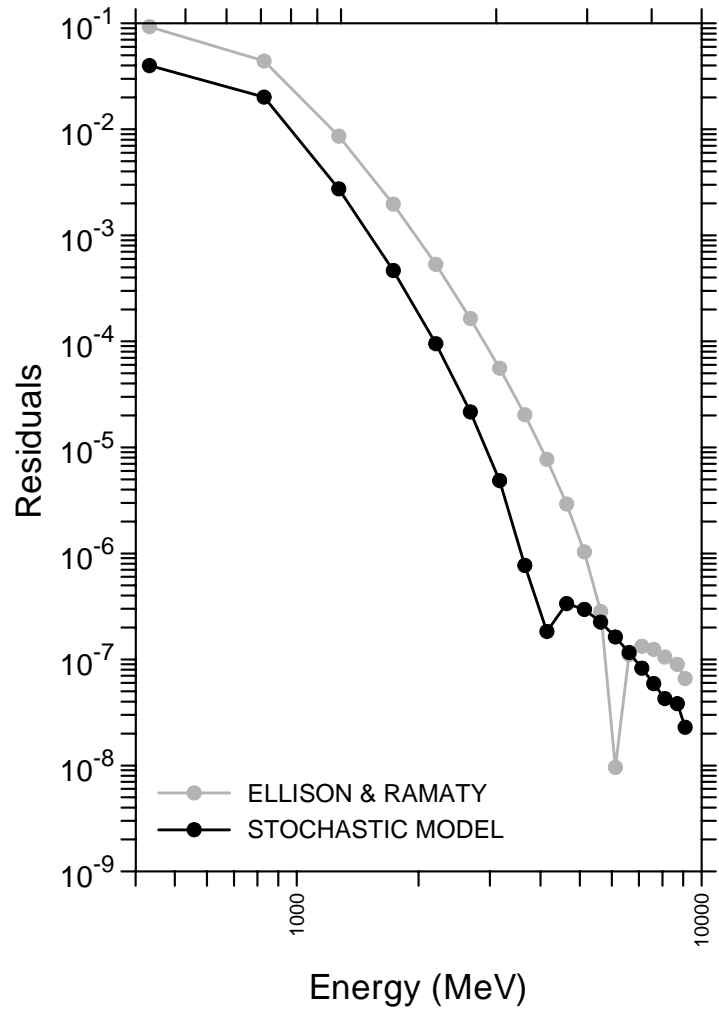

FIGURE 8

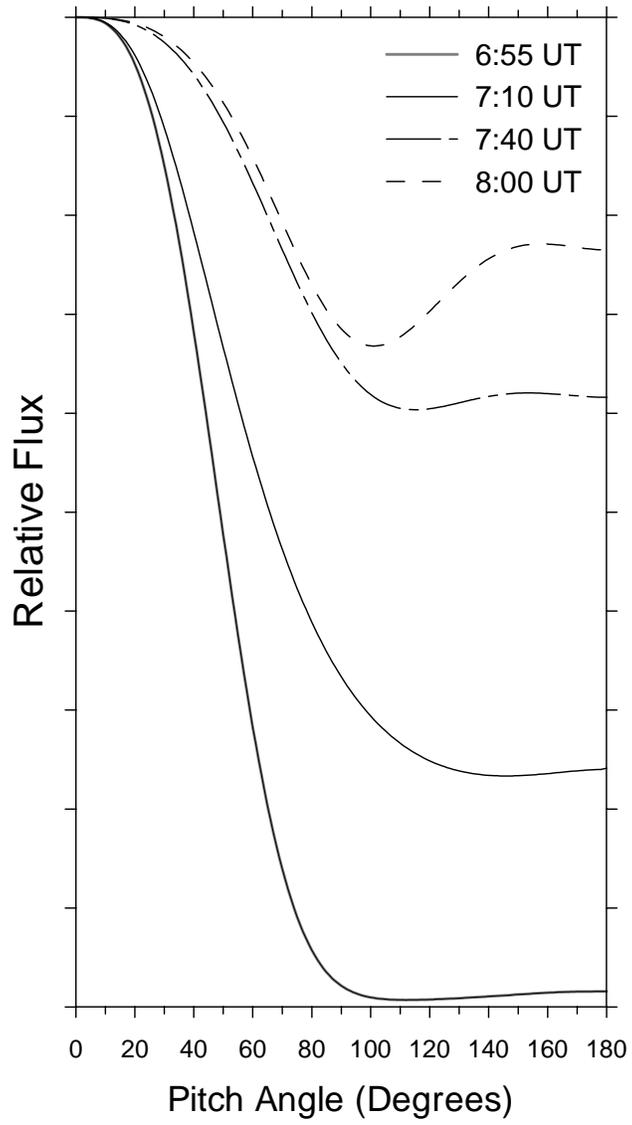

FIGURE 9

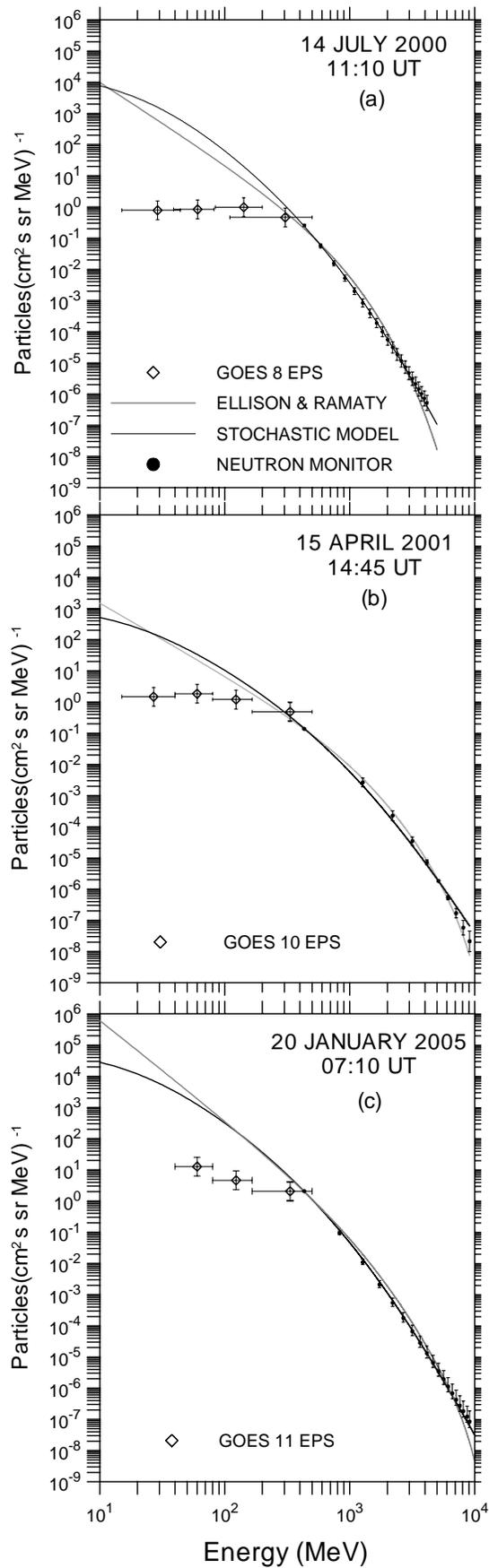

FIGURE 10